\documentclass[10pt,journal,compsoc]{IEEEtran}
\ifCLASSOPTIONcompsoc
  \usepackage[nocompress]{cite}
\else
  \usepackage{cite}
\fi
\ifCLASSINFOpdf
\else
\fi
\usepackage{xcolor}
\usepackage{hyperref}

\usepackage{todonotes}
\newcounter{todocounter}

\usepackage{tabularx}
\usepackage{rotating}
\usepackage{longtable} 
\usepackage{pdflscape} 
\usepackage{array}

\newcommand{\dx}[1]{\textbf{#1}}

\newcommand{\framework}{DX Framework} 

\begin{document}
%
\title{An Actionable Framework for Understanding and Improving Developer Experience}
%
%
%
%

\author{Michaela~Greiler,
        Margaret-Anne~Storey, 
        and~Abi~Noda
\IEEEcompsocitemizethanks{

\IEEEcompsocthanksitem Michaela Greiler is with University of Zurich, Switzerland, and DX, USA.\protect\\
E-mail: greiler@ifi.uzh.ch
\IEEEcompsocthanksitem Margaret-Anne Storey is with University of Victoria, Canada.
\protect\\
E-mail: mstorey@uvic.ca
\IEEEcompsocthanksitem Abi Noda is with DX, USA.
\protect\\
E-mail: abinoda@getdx.com
}
\thanks{Manuscript received August 2, 2021}}

%
%

\markboth{Journal of Transaction on Software Engineering,~Vol.~X, No.~X, October~2021}%
{Shell \MakeLowercase{\textit{et al.}}: Bare Demo of IEEEtran.cls for Computer Society Journals}
%



\IEEEtitleabstractindextext{%
\begin{abstract}
Developer experience is an important concern for software organizations as enhancing developer experience improves productivity, satisfaction, engagement and retention. We set out to understand what affects developer experience through semi-structured interviews with 21 developers from industry, which we transcribed and iteratively coded. Our findings elucidate factors that affect developer experience and characteristics that influence their respective importance to individual developers. We also identify strategies employed by individuals and teams to improve developer experience and the barriers that stand in their way. Lastly, we describe the coping mechanisms of developers when developer experience cannot be sufficiently improved. Our findings result in the \framework, an actionable conceptual framework for understanding and improving developer experience. The \framework\ provides a go-to reference for organizations that want to enable more productive and effective work environments for their developers.
\end{abstract}

}

\maketitle

\IEEEdisplaynontitleabstractindextext

%
\IEEEpeerreviewmaketitle

\section{Introduction}

Improving developer happiness and satisfaction is an important goal for many companies as doing so may lead to higher levels of productivity~\cite{wagnerSystematicReviewProductivity2018, 
graziotinAreHappyDevelopers2013, storeyTheorySoftwareDeveloper2019, 
murphy-hillWhatPredictsSoftware2019} and improve worker 
retention~\cite{westlundRETAININGTALENTASSESSING2008}. 
Previous research has focused on understanding and eliciting factors that help 
describe or predict developer productivity and satisfaction \cite{wagnerSystematicReviewProductivity2018, graziotinAreHappyDevelopers2013, storeyTheorySoftwareDeveloper2019, murphy-hillWhatPredictsSoftware2019, fagerholmDeveloperExperienceConcept2012} 
and which factors may impact developers' productivity and satisfaction with their work
 depends on individual, team, organization 
and project context~\cite{storeyTheorySoftwareDeveloper2019}. 

Developer experience, as defined by Fagerholm and 
M\"{u}nch~\cite{fagerholmDeveloperExperienceConcept2012}, is a broader concept that captures how developers \emph{feel about, 
think about and value their work}. This definition emerged from a review of the literature and these authors claim that developer experience is also shaped by many factors, including team culture, working environment and work activities, but, like satisfaction, is also highly personal.
%
Fagerholm and M\"{u}nch's research provides an initial framework for thinking about the different dimensions of developer experience, but it falls short at identifying which actionable factors impact developer experience, as well as why certain factors may be particularly important and how challenges that impede developer experience may be overcome.  
Notably, some factors and barriers that impact experience may be more \emph{actionable} or easier to act on than others.  
For example, poor directional 
clarity in a project may be easier to act on, than a developer's personality style.

In our research, we build on this previous research to develop an actionable conceptual framework (\framework) that can be used to understand and to guide developer experience improvements. The \framework\ is grounded in empirical data. 
We conducted interviews with a diverse set of software developers across the 
software industry to identify new emergent factors or confirm which of the 
existing factors reported in the literature are perceived as important to them 
(and why) and how they may be actionable at the individual, team and organization levels. 
Through our 
interviews, we also identified strategies developers use to overcome the 
barriers they face, as well as the coping mechanisms they turn to when general improvement 
strategies do not work. Our research has produced a framework that: 
\begin{itemize}
\item outlines \textbf{actionable factors} that may impact developer experience, 
\item elucidates \textbf{contextual characteristics} that make certain factors more or less impactful to a developer's experience
\item identifies \textbf{barriers} that crosscut the factors and that hinder developers 
to improve their developer experience, and 
\item documents \textbf{strategies} and \textbf{coping mechanisms} that developers use to 
overcome these barriers and improve one or more dimensions of their experience as a developer 
on a team. 
\end{itemize}

We anticipate that the \framework\ will help both researchers and practitioners 
understand the most important factors, identify barriers, and implement strategies 
for improving developer experience. 

The paper is organized as follows. In Section \ref{sec:background} 
(\nameref{sec:background}), we 
describe related research on the concept of developer experience 
and other research that reveals many factors that impact different aspects of developer experience. 
We present our study
methodology in Section \ref{sec:method}, describing the interviews and how we analyzed our interview data.  
Our findings lead to the \framework\ that is introduced in Section \ref{sec:framework}.  The factors and why they are important are presented in Section 
\ref{sec:understand} (\nameref{sec:understand}) and the barriers, strategies and 
coping mechanisms that emerged are presented in Section \ref{sec:improve} 
(\nameref{sec:improve}). 
We describe how the \framework\ can be put to use by practitioners and researchers in 
Section \ref{sec:implications}. Finally,  we detail the limitations of our 
research in Section \ref{sec:threats} and conclude the paper in Section 
\ref{sec:conclusion}.

\section{Background}
\label{sec:background}
Before we can understand and eventually improve developer experience, we need to be clear about what we mean by this term, and also review what we know already about the factors that may influence developer experience. 

\subsection{Defining Developer Experience}

The definition we use for developer experience in our research is: 
\begin{center}
	\emph{
``How developers think about, feel about, and 
value their work.''
}
\end{center}

Our definition is inspired by Fagerholm and M\"{u}nch ~\cite{fagerholmDeveloperExperienceConcept2012} and as such is rooted in social psychology from the concept of the trilogy of the mind ~\cite{hilgardTrilogyMindCognition} where the three main dimensions of human experience are cognition (thinking), affect (emotion) and conation (volition to act). 
Consideration of these three dimensions of experience is important as ``real-world problem solving operates in concert with motivational 
and emotional processes, sometimes harmoniously and sometimes 
discordantly.''~\cite[p.~2]{matthewsTraitsStatesTrilogy2004}. 

The cognitive dimension concerns developers' beliefs, and how they think and 
evaluate their development infrastructure, processes, knowledge and skills. The 
affective dimension of the mind describes developers' emotions and how they feel 
about their work. The conative dimension of experience captures developer 
expectations, motivations and how they see the value of their work 
behaviours (their activities, productivity and contributions). Together these 
dimensions interact and shape the intentions for future work behaviour and 
actions. Maylett and Wride similarly define employee experience as the 
experiences, perceptions and expectations of 
employees~\cite{maylettEmployeeExperienceHow2017}. 

Individual personality and other traits shape the three dimensions of the mind, 
but the three dimensions are also shaped by external and social forces, such as 
the nature of the work, the work environment, and whether one is working as part of a team or 
collective. Factors that influence one dimension will typically also influence 
one or two of the other dimensions. The theory of the trilogy of mind aligns 
with our view of the developer experience and also helps us understand the 
important factors that shape developer experience when working with 
others. 
In the following section, we review the factors that have emerged from related 
work on developer experience.

\subsection{Factors That Impact Developer Experience}

Previous research has aimed to identify factors that may impact developer 
happiness, job satisfaction, developer productivity, and motivation. 
We propose that these 
aspects all relate to one or more dimensions of developer experience. 
    
    Graziotin et al.~\cite{graziotin2017unhappiness} 
investigated developer happiness (an affective state) through a survey and found 
that the top factors associated with unhappiness were in order of 
impact: being stuck in problem solving, feeling time pressure to complete 
work, bad code quality and coding practice, under-performing colleague(s), 
feeling inadequate about work, boring or repetitive tasks, unexplained 
broken code, bad decision-making in their team, imposed limitations due to 
infrastructure, and personal issues that are not related to their work. They 
also found that feelings of happiness correlate with perceived 
productivity~\cite{graziotin2017unhappiness}.
    They found that happier developers tended to perceive higher productivity 
and vice versa. Bellet et al. also report a strong relationship 
between productivity and happiness~\cite{belletDoesEmployeeHappiness2019}.

Murphy-Hill et al. investigated which factors can 
predict software development productivity through a study with three companies 
~\cite{murphy-hillWhatPredictsSoftware2019}. They found that being enthusiastic 
for one’s job was the top predictor for productivity, followed by having 
supportive team members of their ideas and having autonomy over their tools and work~\cite{murphy-hillWhatPredictsSoftware2019}.
A literature review conducted by Wagner et al. details even more factors that have been found to associate with 
reported or perceived productivity. 

Storey et al. researched developer satisfaction with their work building on 
Wright and Cropanzano's~\cite{wright2000} definition of job satisfaction as 
``an internal state that is expressed by affectively and/or cognitively 
evaluating an experienced job with some degree of favor or disfavor''. The 
satisfaction factors Storey et al. identified through a large survey at Microsoft 
include: doing impactful work, being an important contributor on their team, 
having a positive work and team culture, feeling productive, receiving 
appreciation and rewards, and experiencing a positive work-life balance~\cite{storeyTheorySoftwareDeveloper2019}. They 
also found a bidirectional relationship between satisfaction and developer 
productivity, and that there are additional factors that influence productivity: 
autonomy in one's work, ability to complete tasks, and the quality of the 
engineering system. 

Motivation for work aligns most closely with the conative dimension of developer 
experience. Steglich et al.~\cite{steglich_social_2019}, through a literature review, summarized 
a number of social factors that influence developer's motivation to work in mobile software system ecosystems, 
some of which are the opportunity to learn something new, to have ``fun'' and to collaborate with others.  
Beecham et al. also studied developer motivation in a systematic 
literature review, and identified many personal and work characteristics that 
moderate the influence of a large number of motivating factors (e.g., good 
managers, task fit, empowerment) and demotivating factors (e.g., poor managers, 
poor working environment, stress)~\cite{beecham2008}. 
Subsequent work by Sharp et al.~\cite{sharp2009} 
proposed a model of motivation in software engineering that includes motivators, 
outcomes, characteristics and context. More recently, empirical studies by 
Fran\c{c}a and colleagues~\cite{franca2012} identified a variety of 
factors that affect motivation, such as career progression and autonomy. Fran\c{c}a 
et al. point out that motivation and job satisfaction are not the same thing, which 
aligns with the three-dimensional view of developer 
experience proposed by Fagerholm and 
M\"{u}nch~\cite{fagerholmDeveloperExperienceConcept2012}, with motivation more closely 
aligned with the conative dimension, and satisfaction with the cognitive dimension.

Previous research has collectively identified hundreds of factors, and 
many factors overlap because developer satisfaction, productivity and motivation are related. 
Our research spans not just one or two constructs, such as satisfaction or productivity, but factors which affect developer experience overall. Hereby, 
our focus is to identify which of these factors may be the most important, why they are important and which ones are actionable. 
In the next section, we describe our research method for identifying these actionable factors.


\section{Methodology}
\label{sec:method}

Our methodology involved semi-structured interviews with a diverse 
set of developers in terms of role, industry, projects and experience. 
In this section, we 
describe how we selected interview participants, the interview questions we 
asked, and our approach for analyzing the results.

\subsection{Research Questions}
\label{sec:research-questions}
Congruent with our qualitative research approach, our research questions were 
emergent and refined as we gathered and analyzed our data. Our initial guiding 
research question was: what are the most important and actionable factors that 
affect developer experience?  As we conducted our interviews, we not only 
uncovered factors that matter to developers, but also found contextual characteristics of 
their work which determine their relative importance to developers (e.g., developer seniority). Additionally, 
respondents shared barriers impeding their ability to improve developer 
experience, strategies they and their teams employed to successfully make 
improvements, and their coping mechanisms when factors negatively impacting 
developer experience were not improved. Our emergent research questions were as 
follows: 

\begin{itemize}
\item RQ1: What \textbf{important factors} affect developer experience?
\item RQ2: What \textbf{contextual characteristics} influence how important a factor 
is to a developer's experience?
\item RQ3: What \textbf{barriers} impede developers and their teams from 
improving factors that affect developer experience?
\item RQ4: What \textbf{strategies} do developers and their teams employ to 
improve developer experience?
\item RQ5: What \textbf{coping mechanisms} do developers resort to when factors 
that negatively impact developer experience are not improved?
\end{itemize}

\subsection{Semi-Structured Interviews}
Through semi-structured interviews with 21 software developers, we set out to 
explore what factors developers perceive as affecting developer experience 
and to understand how they improve these factors on their teams. Each interview 
took between 45 and 90 minutes. We used Zoom to record each interview and 
transcription software to transcribe each recording. 
Although each video recording was automatically transcribed, the researcher coding the interviews always watched the original video during the coding process. That way, we ensured the best and most accurate interpretation of the interview possible. The transcripts were mainly used to search within the interviews and be able to make more connection between the memos, codes and concepts. We also corrected any mistakes that we saw in the automatic generated transcripts during the coding process. In order to make this process efficient, we used timestamps to make sure we can easily retrieve relevant parts of the video recordings, without the need of watching the whole video over and over again. This means that each memo, code or concept had a pointer to the videos this insight came from, as well as the start and the end time of relevant segments of the videos.

The interview questions were based on an interview guide which can be found at 
\url{https://github.com/get-dx/dx-framework}. 
In the first part of the interview, we provided each participant 
with a high-level definition of developer experience: ``developer experience is 
the developer's perception of the work, processes, and culture that they 
encounter while building software on a team''.\footnote{Note this preliminary 
definition of developer experience was used to initiate the discussion in the 
interviews, rather than a formal definition as we provide in Section 
\ref{sec:background} to frame our research.} We then asked them which factors they 
perceived as affecting their experience.

In the second part of the interviews, we guided participants through a 
discussion on factor importance and actionability. To deepen the discussion, we 
showed participants a list of factors that we found in related literature. These 
factors served as a prompt to help deepen the discussion and encourage 
participants to consider factors that were not immediately top of mind. This 
list was assembled by consolidating factors from the literature we presented in 
Section \ref{sec:background}, merging duplicate factors, and then reducing the 
list to factors we considered actionable for developers and their teams. We 
also grouped the factors into categories to present to the participants in order to reduce cognitive overload in the interviews. This list of categorized factors is available in our supplementary materials found at \url{https://github.com/get-dx/dx-framework}.  

In the third part of the interviews, we focused on understanding whether and how 
participants could influence or improve their developer experience. Based on the 
insights and experiences shared by each participant, we adjusted and refined our 
questions about importance and actionability to match the circumstances of the 
interviewee. This allowed us to investigate uncovered areas and continuously 
gather more perspectives on topics that previous participants introduced. 

\subsubsection{Interview Participants}
To select participants for this study, we used convenience sampling by reaching 
out to developers in our network using email or other social and communication channels. 
Our selection criteria was that participants had to have more than six 
months of professional software development experience and currently be employed 
as a developer or development lead. Prior to each meeting, we asked participants 
for their consent to be interviewed and asked for their permission to record the 
session. We informed participants that they could withdraw from the interview at 
any time and that their responses would then be deleted. Our consent form and 
interview instructions can be found in our supplementary materials online. 

We aimed for a diverse set of participants, in terms of experience, companies, countries and gender. We were able to recruit participants from various countries and regions within America and Europe. Although, we also aimed for including more women, only one agreed to be interviewed.

16 of the 21 participants had more than six years of professional software 
development experience and five of those had 20+ years of experience. Four of the
participants had between two and five years of experience and one participant 
had six months experience as a professional developer. 
To make sure we included junior developers with less than five years of experience, we 
recruited participants from developer communities that we belong to.
Participants worked in a variety of industries (including the medical sector, 
developer tooling, human resource software, and general consulting). Our participants' team sizes varied 
from 2 to 100 people, and their company sizes varied between 5 and 20K+ people. 
Table~\ref{tab:participants} shows a summary of the developers we interviewed.

\begin{table}
\centering
\begin{center}
\begin{tabular}{ |p{1em}|p{3.7em}|p{5.5em}|p{2em}|p{5.5em}|p{4.3em}| }
	\hline
 No. & Company Size & Industry & Team Size & Current Role & Experience \\ 
 \hline
 P1 & $\sim$2300 & Developer Tools & 7 & Lead\linebreak Engineer & 22 yrs \\    
 P2 & $\sim$1500 & Payroll \& Human Resources Software & 8 & Tech Lead & 6.5 yrs \\  
 P3 & $\sim$80 & Medical\linebreak Sector & 5 & Tech Lead & 14 yrs \\  
 P4  &  $\sim$1300 & Software &  & Team Lead \& Engineer & 20 yrs \\  
 P5 & $\sim$500+ & Energy & 10 & Solution\linebreak Architect & 7 yrs \\  
 P6 & $\sim$300 & Software & 100 & Senior Software\linebreak Developer & 20+ yrs \\  
 P7 & $\sim$80 & Health care & 15 & Fullstack Developer & 4 yrs \\  
 P8 & $\sim$20K+ & Commerce & 50 & Software Developer & 4 yrs \\  
 P9 & $\sim$5 & CRM Software & 3 & CTO/Tech Lead & 15 yrs \\  
 P10 & $\sim$1200 & Software Industry & 8 & Senior Fullstack\linebreak Engineer & 8 yrs\\
 P11 & $\sim$20 & Consulting & 20 & Software Developer 3 & 5 yrs\\
 P12 & $\sim$180 & Video Streaming & 8 & Engineering Director & 23 yrs\\
 P13 & $\sim$150 & Education & 35 & Team Lead \& Engineer & 4.5 yrs\\
 P14 & $\sim$180 & Video Streaming & 8 & Software Developer & 6 yrs\\
 P15 & $\sim$50 & Legal Tech & 5 & Senior Software\linebreak Engineer & 6 yrs\\
 P16 & $\sim$20 & Software & 6 & Software Engineering Intern & 6 mths \\  
 P17 & $\sim$125 & Software Consulting & 8 & Software Developer & 2.5 yrs\\
 P18 & $\sim$150 & Education & 2 & Junior Software\linebreak Engineer & 4.5 yrs \\  
 P19 & $\sim$80 & Human Resources \& Recruiting & 10 & Developer & 9 yrs \\  
 P20 & $\sim$20 & Software Communications & 4 & Staff Engineer & 16 yrs\\
 P21 & $\sim$8 & Software & 3 & Staff Fullstack Engineer & 23 yrs\\
	\hline  
\end{tabular}
\caption{Details of Study Participants}
\label{tab:participants}
\end{center}

\end{table}

\subsubsection{Interview Process}

We conducted two pilot studies before finalizing our interview guide and 
conducting the final set of 21 interviews (the pilot study data was not used in 
our analysis). The pilot studies encouraged us to add a definition of developer experience (listed above) 
to the interview protocol, and to include a list of factors distilled from the literature 
that may impact experience as a prompt 
to encourage more discussion (which was needed for some participants more than 
others). As we describe in the following section, we iteratively coded the data 
from the interviews and stopped conducting additional interviews once we 
determined that our codes and insights were fully saturated. As  no new 
insights or codes emerged from the three latest interviews, we stopped recruiting further participants.

\subsection{Coding Process and Developer Experience}

To analyze the interviews, we used an open coding approach where we coded the 
interviews in an inductive (bottom-up) way~\cite{charmaz2006constructing}. 
Interviews were conducted and coded by two or more authors over several 
iterative cycles. Interview recordings and transcriptions were continually 
revisited in an iterative process. For each new interview, 
we went back to previous interviews to see if previous interviewees also mentioned
the new insights. As mentioned above, once no new codes and insights emerged in three consecutive interviews, we concluded our findings were saturated and stopped recruiting new participants.

We divided the transcripts of the participants into coherent units 
(sentences or paragraphs) and added \textbf{preliminary codes} that represented 
the key points that each participant talked about. We later agreed on a 
set of \textbf{focused codes} that captured the most frequent and relevant 
factors of developer experience.   We then used \emph{axial coding} as described by Charmaz~\cite{charmaz2006constructing} to group the codes 
into \textbf{categories}. This was done using visual mapping tools in several 
iterative cycles with discussions among the authors. As we were coding, we wrote 
memos for the codes and categories, and noted relationships across codes. Table 
\ref{tab:codes-example} shows examples of the coding process for several transcripts and the resulting codes, categories and core categories (that aggregate categories in our code hierarchy).
    
    Early in our analysis, we identified a number of emergent \textbf{core 
categories}: developer experience (DX) factors, contextual characteristics that may impact the importance of 
factors, barriers impeding development teams from improving developer experience,
strategies for improving DX factors, and coping mechanisms developers resorted to if they were not able to improve their experience. These five core categories are key components in the \framework\ that emerged from our analysis (as shown in 
Figure~\ref{fig-framework}). The \framework\ is our main research outcome and 
it also helped us refine our preliminary research questions (as presented above in 
Section~\ref{sec:research-questions}). The core categories and associated 
subcategories and codes will be described in more detail in the following 
sections.

When we started our analysis, we anticipated that we might be able to align the factors that would emerge from our study with the three dimensions of the trilogy of mind but we did not find this. 
For example, one of the factors that emerged from our study was codebase health. Codebase health impacts how developers think about their work, but it also impacts how they feel about their work (e.g., frustrated, proud) 
and their motivation to work. 
In hindsight, this makes sense, as Matthews et al. mention
``it is necessary to treat cognition, emotion, and motivation as inextricably related''~\cite{dai2004motivation}.

\begin{table*}
\centering
\begin{center}
\begin{tabular}{
>{\raggedright}p{0.3\textwidth}|
>{\raggedright}p{0.2\textwidth}| 
>{\raggedright}p{0.15\textwidth}|
p{0.15\textwidth}}
\hline
\hline 
\multicolumn{4}{p{0.9\textwidth}}{
\textbf{Transcript Unit}:  If you have to touch a certain piece of code every other week and the whole codebase around it is suboptimal and flawed, then making changes to that code is always very difficult and a little dangerous. Or if it's not tested well, or if you don't even understand the full scope, because there's so many things that are attached, then making changes, even for new features to that part of code is not a nice task, I'd say. (P3)}\\
\hline
 \textbf{Preliminary Code:}\linebreak Changes can be difficult in low-quality codebase, high risk making changes 
 & \textbf{Focused Code:}\linebreak Codebase health 
 & \textbf{Category:}\linebreak Development and Release
 & \textbf{Core \linebreak Category:}\linebreak DX Factor\\
\hline 
\hline
\multicolumn{4}{p{0.9\textwidth}}{
 \textbf{Transcript Unit}:
 "the developer tools that people use [are affecting developer experience]. 
 So,  I work in the.net stack mostly,  and the tooling is actually great. But there is a tool called ReSharper,  which is like an add-on that companies have to pay for,  which makes you a lot more productive. And other types of tools. So I think tooling and the development environment itself is another big piece of it." (P22)}\\ 
\hline
 \textbf{Preliminary Code:}\linebreak Developer tools influence productivity
 & \textbf{Focused Code:}\linebreak Development environment
 & \textbf{Category:}\linebreak Development and release
 & \textbf{Core \linebreak Category:}\linebreak DX Factor\\
 \hline
 \hline
\multicolumn{4}{p{0.9\textwidth}}{
\textbf{Transcript Unit}: The founders of the company, when we would go to them for that criticism, 
or what we need - they weren't very responsive for it. They didn't care, or at 
least they didn't show any sort of empathy about it or any understanding of our 
situation. They just said keep working. And so, I think there was always effort 
to try to improve, but we started noticing the patterns and gave up trying. We 
knew that at some point it doesn't matter. We just need to do our job. And that 
was kind of the reason why we would just complain to each other because there 
was no point to reach out anymore because we tried and it didn't work. (P11)
}\\ 
\hline
\textbf{Preliminary Code:}\linebreak Developers stop speaking up when not heard 
& \textbf{Focused Code:}\linebreak Stop speaking up 
& \textbf{Category:}\linebreak - 
& \textbf{Core \linebreak Category:}\linebreak Coping Mechanism \\

\hline
\hline
\end{tabular}
\caption{Illustration of the coding process. More examples of the coding process can be found as part of our supplemental material.}
\label{tab:codes-example}
\end{center}

\end{table*}

\section{The DX Framework}
\label{sec:framework}

\begin{figure*}
\includegraphics[width=1\textwidth]{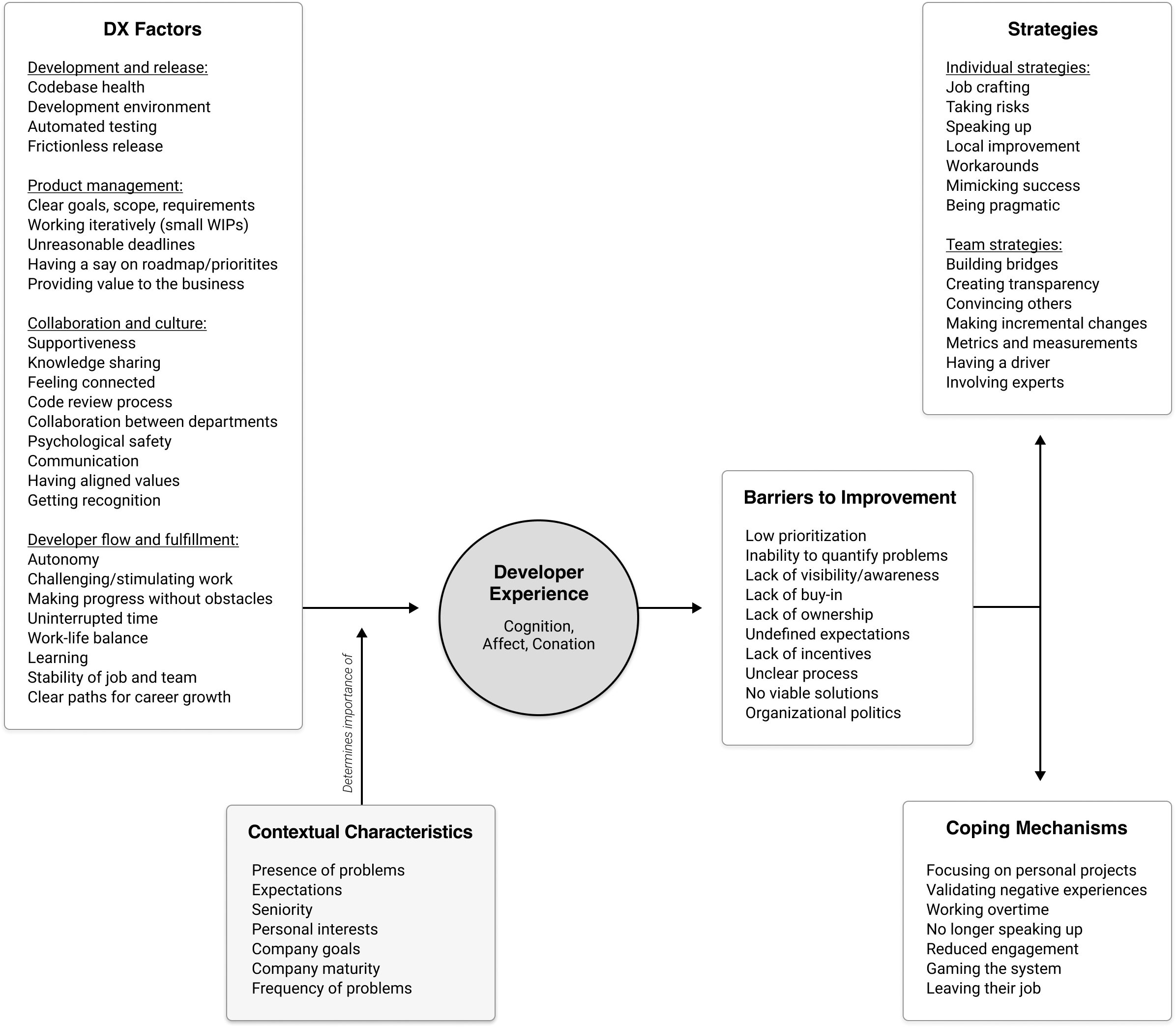} 
\caption{DX Framework.  A larger version can be found at \url{https://github.com/get-dx/dx-framework}.}
\label{fig-framework}
\end{figure*}

The main outcome from our research is the ``\framework'' (see Figure~\ref{fig-framework})\footnote{A larger version can be found online at: \url{https://github.com/get-dx/dx-framework/blob/main/DX-Framework.jpg}}. The central concept in our 
framework is \textit{\textbf{Developer Experience}} which is characterized by the trilogy of 
mind dimensions (cognition, affect, and conation). This central concept 
in the \framework\ is inspired by other research (as discussed in Section 2). The 
other two parts of the \framework\ (the left and right sides) emerged from our 
research. 
      
On the left side of the \framework, we list the two core categories that emerged from our research that relate to \textit{\textbf{understanding developer experience}}. 
These two categories include the factors that emerged from the first 
part of our interviews (the factors that developers shared impacted their experience, without any prompts) and contextual characteristics that moderate factor importance to the participants. We describe 
these core categories in Section~\ref{sec:understand}. 
     
On the right side of the \framework, we show the three core categories that 
emerged from our research that are concerned with \textit{\textbf{improving 
developer experience}} and include the barriers to improving developer 
experience, the strategies for improving developer experience, and the coping mechanisms developers resort to when they cannot improve their experience.  We discuss these core categories in Section~\ref{sec:improve}.

\section{Understanding Developer Experience}
\label{sec:understand}

\subsection{Factors Affecting Developer Experience (DX Factors)}

We aimed to identify what developers perceive as the most important factors affecting developer experience. 
As described in Section~\ref{sec:method}, we iteratively 
coded their responses to the question ``what factors affect your experience'' and 
grouped them into categories. These categories represent themes that helped us 
understand a set of factors as a group. Note that the factors we share 
emerged from our interviews before we prompted the participants to 
consider other factors that we had listed from the literature. That means, that we do not include additional factors that were only mentioned or discussed after participants had seen our list of potential factors. 
This list was only used to steer the discussion around
importance and actions that developers take or do not take to improve developer experience. 
We discuss the emerged DX factors, our 
focused codes for this core category, and the categories we grouped them into 
(through axial coding) in the following.

 We note that the categories we assigned to the factors 
 are based on our own experiences and knowledge of software 
development, and that other categories could also be used for describing these 
important factors. We do not count how many times each factor emerged as the 
interviews were open ended, so any counting we would do could be misleading. 
However, the factors that emerged were all mentioned by two or more 
participants. We do preserve a mapping of participants to factors in our raw 
data that is available online for research transparency purposes. 
Throughout the following, we also discuss how other researchers have similarly identified these factors impacting some aspects of developer experience.  

\subsubsection{Development and release}

The development and release category consists of factors relating to developers' 
codebases, as well as the tools used to write and release code.

\dx{Codebase health} refers to the quality, 
maintainability, and ease of working in a codebase and its impact on developer experience. As P6 shared: \emph{``Part of 
[what affects developer experience] is the codebase itself. This is a tough problem, right? 
Working on a legacy codebase. It was poorly architected and just hard to 
understand. Tightly coupled. All the things that make a codebase tough to work 
with and tough to change. That is actually a major factor.''} 

The \dx{development environment} was an additional factor. P1 shared: \emph{``the very first thing that came to my mind was 
around, the tooling or any friction around the tooling that either makes it 
really painless to go from I'm working on an idea to I'm testing that idea and 
production, the tooling that makes it painful to go from point A to point B''}.  
They also added: \emph{``[what affects developer experience is] how quickly I can compile my code, how fast continuous integration 
runs, how long it takes me to be able to deploy my change to a lab environment 
or production environment, how reliable my tests are [...] 
anything that extends the feedback loop.''} This factor also includes aspects of engineering infrastructure such as the setup and configuration, as well as debugging and monitoring. 

Another factor participants shared was having sufficient 
\dx{automated testing}. P12 stated: \emph{``insufficient test 
coverage, for example, which makes it incredibly hard to do any changes, and 
then also super complicated codebases where things are really badly designed. 
So, really it's about confidence in that case. Everything that builds 
confidence, that makes it easy to understand what's going on and to be  confident 
that the change you're doing is the right one. That is a big thing for developer 
experience to me.''}

Lastly, a factor that emerged in several interviews was \dx{frictionless releases}. For example, P12 stated: \emph{``We spend a lot of effort on reducing cycle time, on improving the experience for PRs, to reduce the time it takes get things out in production and get them safely out into production. So all these things come together to deliver more software and to be able to do small, incremental improvements without experiencing any problems.''}

Previous research has also identified that the above factors relating to 
engineering systems (codebase, tools and processes) have an impact on developer 
productivity and satisfaction~\cite{storeyTheorySoftwareDeveloper2019, murphy-hillWhatPredictsSoftware2019, besker_influence_2020}.

\subsubsection{Product management}

Having \dx{clear goals}, 
\dx{scope} and \dx{requirements} can greatly affect developer experience.
P4 described how extra effort 
is needed to clarify with others that they are doing the right tasks, and that a 
large task may need to be divided into manageable tasks (what Schmidt and Bannon describe as articulation work~\cite{schmidt1992taking}): \emph{``On the process side, [developer experience] also means that 
clear goals really help me. If there's lots of uncertainty around the tasks, or 
if it's like very large tasks that are ill-defined, then it feels a bit like 
walking in a swamp. It's very hard to make progress. And just the clarification 
work you have to do to break down tasks or to get more certainty for tasks, 
doesn't feel like actual work. For development teams, it doesn't feel like it, 
because you don't submit a PR [instead] you talk to lots of people and write a 
document and then you can basically create five tasks from one. And that doesn't 
feel like you achieved something because you're not deploying, you're not 
writing code.''} P4 also touched on \dx{working 
iteratively}, in small tasks, which many participants described as a DX factor.

\dx{Unreasonable deadlines} causing poor tradeoffs in terms of 
timeline, roadmap, and priorities were things that many participants described 
as stressful and negatively impacting developer experience. P13 stated: \emph{``The 
goals and ambitions that the product team wants - I understand them and I 
see that these projects are important, but I think that the timeframe that they 
want them in is not conducive to developing good software and lasting software. 
There's sort of a struggle.''} P13 also added: \emph{``I sort of have seen that 
high stress projects with tight deadlines can lead to definitely degraded 
experience, and interpersonal interactions between engineers.''}  
The impact of 
poor timelines on developer happiness was also reported by Fagerholm et al. \cite{graziotin2017unhappiness}.
In addition, participants talked about how 
\dx{having a say on roadmap/priorities} improves developer 
experience.

Many interview participants talked about how \dx{providing value to the 
business} was important to them. For example, P11 said: \emph{``For me, 
what I find the most valuable in my work is that I'm providing value to the 
business. So, if I'm able to find a bug that may not be fun or interesting - as 
long as it unblocks the business, I feel that I've done something rewarding that 
day.''} Previous research has also shown that
doing work that feels impactful is important to developer 
productivity and satisfaction 
\cite{storeyTheorySoftwareDeveloper2019,murphy-hillWhatPredictsSoftware2019}.

Participants also mentioned how unforeseen changes in product 
direction can lead to ``thrown away'' effort. As P20 explained: \emph{``The absolute 
worst developer experience is when I feel like I have a very clear goal and I 
get done or almost done, and it's like, oh no, we have to change it completely 
because then you have to start from the beginning.''} P7 talked about how 
agile processes lead to changing requirements: \emph{``A lot of 
work that I do ends up being replaced by some other work, like immediately. I 
mean, I see the results of my work, but maybe 30\% of it goes to nothing exactly 
because the situation changed.''}

\subsubsection{Collaboration and culture}
The collaboration and culture category includes factors that relate to the 
relationships people form and how those hinder or help them complete their work.

One frequently mentioned factor was \dx{supportiveness}---developers rely on   
support from their peers and getting help quickly when 
they are stuck. P17 explained: \emph{``I would say one of the biggest ones is 
the amount of time that more experienced developers have to spend with you. I 
noticed that on days where everyone's busy [...] and people don't have time for 
me, my frustration level goes up exponentially. Where I'm struggling with a 
problem for hours that I know someone already has the answer to, but they don't 
have time to give me a little bit of guidance first [...] That's probably the 
single biggest factor to whether I have a good or bad day at work.''}
The impact of supportive team members on team culture and productivity has been reported in previous research, for example, Schneider et al. present how positive comments in meetings positively impacts team behaviour~\cite{schneider2018positive}, while ``supporting of new ideas'' was one of the top factors for predicting productivity as reported by Murphy-Hill et al.~\cite{murphy-hillWhatPredictsSoftware2019}.

We found that junior engineers and new team members especially rely on the 
support of their peers. But more experienced engineers also talked about the importance 
of \dx{knowledge sharing} and \dx{feeling connected}. P14 explained: 
\emph{``There is this connectedness to the 
colleagues around you and it just helps if you have a very direct and 
instantaneous communication with your peers around you. Which really helps you 
build connections,  which also helps in how you support each other in a team. 
And I think that's very important for a developer because I am a full stack 
developer, so we touch several systems with several programming languages, and I 
cannot be aware of every single system. So in some sense, I really rely on 
information from my colleagues. So if I have this spontaneous connection to my 
colleagues around me, it is much easier to  get information and just understand 
and communicate what needs to happen in certain systems.''} 
Feeling connected and the flow of information also emerged as critical factors for team productivity in a recent study of developers working from home during the pandemic~\cite{miller2021your}. 

\dx{Code review process} was a distinct factor related to knowledge sharing and feeling supported. 
Participants shared how code reviews contribute to a better codebase as well as help 
facilitate knowledge-sharing and mentorship. However, poor code review process can also 
hurt developer experience. For example, participants (P10, P16) talked about how `nitpicking' 
or being overly critical about their changes during code reviews negatively 
impacted their developer experience. And P1 talked about how the feedback tone 
in code reviews is critical. 
The important role of code review in positive team culture is also discussed by Bacchelli et al.~\cite{bacchelli2013expectations}.

Good collaboration is not just important within teams---nearly 
all participants talked about how \dx{collaboration between 
departments} is also important for developer experience. Developers 
specifically called out collaborations with product management, 
design, and quality assurance teams. 

\dx{Psychological safety}, i.e., feeling safe and being able to speak one's  
mind is an important factor affecting developer experience. Several 
participants talked about how having an open-minded culture where less 
experienced developers can speak up was important to them, and how leadership 
plays an important role in ensuring psychological safety.
Lenberg et al. shared how psychological safety helps predict higher self-assessed performance and job satisfaction~\cite{lenberg2018psychological}. However, 
their research shows clarity of norms is a stronger predictor, 
which relates to our product management factor of having clear goals and expectations. 

Other important factors are \dx{communication} between developers and 
teams, \dx{having aligned 
values}, and \dx{getting recognition} for work from peers and managers. These factors also emerged in other research on developer satisfaction, happiness and productivity~\cite{graziotin2017unhappiness, murphy-hillWhatPredictsSoftware2019, storeyTheorySoftwareDeveloper2019}.

\subsubsection{Developer flow and fulfillment}
Participants talked about 
factors that influence the degree to which thay can perform their tasks with ease and joy, as well as how they perceive their future at the company. We grouped these factors into the category of developer flow and fulfillment.

One frequently mentioned factor that influences developer experience is \dx{autonomy}. P3 
described: \emph{``How much freedom I have in making technical decisions versus 
somebody telling me which approach to take for a certain problem.''}
But not every developer wants total autonomy. Several participants 
talked about how having autonomy is important, but that it must be bounded. 
For example, P5 explained: \emph{``Autonomy. That's one of the 
things that comes to my mind, but not let's say unbounded autonomy. But kind of, 
know your limits in a way. Know where you can go towards and where you can't. So 
you know, where it's safe for you to  experiment and be creative about, and so 
on. [...] One of my struggles today is not knowing where I cannot go to. Not 
knowing what's off limits and that means that anything can be off limits. So you 
kind of feel trapped in a way.''} The importance of autonomy on developer satisfaction, happiness and productivity has emerged in other research as well ~\cite{graziotinAreHappyDevelopers2013, murphy-hillWhatPredictsSoftware2019, storeyTheorySoftwareDeveloper2019}.

Participants talked about having optimally \dx{challenging and stimulating 
work}, where work is not boring but also not overwhelmingly difficult. 
P20 explained: \emph{``The other part of [developer experience] is the work 
that I'm doing in and of itself. Like, there is kind of like those bands: 
[starting with] `this is just easy, busy work' to there's that middle ground of 
`I'm learning something new, this is challenging, but I'm not in over my head', and 
then like kind of you've got that top level of `here do this, and you have no 
idea what you're doing'. So, like having the correct, I guess, level of work.''}  
What participants described is what Csikszentmihalyi~\cite{csikszentmihalyi2014flow}, one of the co-founders of 
positive psychology, described as one of the ingredients to get into a flow-like 
state: a balance between challenge and skills. Flow is also seen as a key dimension for developer productivity~\cite{forsgren2021space}. 

Other factors included in this category that relate to the joy and flow 
of work are \dx{making progress without obstacles}, having enough 
\dx{uninterrupted time}, \dx{work-life balance}, and the degree to 
which developers can \dx{learn} on the job. Lastly, participants shared 
the \dx{stability of their} \dx{job and team} and having  
\dx{clear paths for career growth} as factors affecting developer 
experience.

\subsection{Factor Importance}

During the second part of our interviews, we asked participants to describe why certain factors were important to them.
To guide this discussion, we also showed them the list of factors we derived from literature to prompt them to consider factors that were not immediately top-of-mind but might still be important to them. 

\textbf{Psychological safety is paramount.}
One consistent pattern we observed is that participants described \emph{psychological safety} as 
more important than other factors. 
P9 said: \emph{``I don't think that a not perfect codebase 
is a reason for somebody to leave the company. Let's put it that way. 
Whereas if you have a toxic culture in your company, that's when people think of 
leaving a company and actually really do it.''}
P12 expressed in similar terms: \emph{``Culture is a very big 
thing. I would even say that these are like baseline traits that you have to 
have otherwise, everything else really does not matter.''}

Other participants explained that 
good culture is crucial to enable improving all other factors and developer
experience overall. 
They mentioned how poor  
psychological safety discourages developers from speaking up about problems or making proactive improvements. 
They also shared that on teams with good psychological safety and culture, 
developers are more willing to 
voice and tackle problems in order to continuously improve developer experience. 

For other factors that were shared through our interviews, 
participants described importance in terms of their personal or project circumstances. 
These contextual characteristics are outlined below and 
illustrated in the \framework\ (see Figure \ref{fig-framework}) where we show these as moderating~\cite{mackinnon_integrating_2011} the impact of factors on developer experience. 

\textbf{Presence of problems trumps other factors.}
One common theme that emerged is that developers perceive the \emph{presence of 
problematic factors} more strongly (and therefore as being more important) than factors which contribute positively to their experience. 
This effect is known in research as the 
positive-negative asymmetry effect or negativity bias~\cite{rozin2001negativity}, which  
describes how negative events have a bigger impact on a person's experience 
(and memory) than positive events (even of the same type). 

\textbf{Expectations determine importance}.
Participants expressed that \emph{problems that are 
expected} or perceived as normal bother them less than problems that they 
envision could be better. For example, P10 
said: \emph{``It's like so normal that the [Continuous Integration], like everything is slow, that it 
is no longer something that is so much discussed. Because, it also takes more 
time and we have a specific team that is just doing that.''} 

\textbf{Seniority influences importance.}
Another common theme mentioned by participants is that \emph{seniority} 
affects how important developers perceive certain factors to be. 
Senior engineers were described as having greater mental capacity to think and 
care about a wider variety of factors, such as release process or team culture, 
whereas junior developers were described as being more focused on developer flow and codebase 
health. 
P2 stated: \emph{``The weight of these 
categories changes drastically depending on how senior an engineer is. So if 
you're a junior developer, you might actually really only care about codebase 
health and the developer workflow.''} Senior developers might also be expected to
be responsible for certain factors, making these factors more important. 


\textbf{Personal interest affects importance.} Participants described 
how personal interests could make factors more or less important. For example, P2
stated: \emph{``People can be so disconnected from some of these other 
areas of building product that it's almost like apathy. Like they don't care 
about the direction of the product. They just have a ticket that they need to 
get done and that they're not really questioning or thinking about the product 
direction, or how the team as a whole can continuously improve. They're just 
trying to get their stuff done.''} 

\textbf{Company goals make factors more or less important.} Company goals affect which factors are perceived as important. 
For example, if a company has a goal of releasing frequently, factors 
associated with release become more important to developers 
because of increased expectations that these factors should be improved. 
P10 shared: \emph{``For delivering fast, it depends on the company goals. If you're a company that releases three to four times a year, the deployment process does not need to be optimized 
that much versus [a company] deploying every day.''}

\textbf{Company maturity defines what is important.} Developers often described how the \emph{maturity of a company} impacts 
factor importance. The more mature a company or team, the higher its expectations are for having a good development environment, or for having dedicated teams that handle design and testing.

\textbf{Frequent problems are more important.}
Lastly, many participants described a relationship between \emph{frequency} and 
importance---the more frequently a problem occurs, the more negative impact it has.

\section{Improving Developer Experience}
\label{sec:improve}
A primary goal of our study was to understand how developers and their teams \emph{can} improve their developer experience. To investigate this, we asked developers about situations where their developer experience was less than ideal, and probed about how they and their teams dealt with it. 
From this data, we learned about \emph{barriers} that developers and their teams face in improving developer experience along with \emph{individual and team strategies} they employ to make improvements. In addition, we identified the \emph{coping mechanisms} that developers use when areas of friction in their developer experience were not improved. The barriers, improvement strategies and coping mechanisms presented in this section are important parts of our emergent \framework.

\subsection{Barriers to Improving Developer Experience}
We asked developers what inhibited them and their teams from improving developer experience. We discovered that the barriers developers face are primarily organizational rather than technical. The list of barriers are summarized on the right-hand side of Fig.~\ref{fig-framework}.

\textbf{Low prioritization.}
One of the most prevalent barriers developers face is getting improvements \emph{prioritized}. Developer experience cannot be improved without committing time and resources, and organizations often prioritize other objectives over developer experience. For example, developers talked about how reducing technical debt was seen as a lower priority than shipping features: \emph{``Although our product manager is a really kind guy and he knows that technical depth needs to improve, we sometimes fight hard to get a slot in our sprint for just improving things [because of] pressure. Pressure from the company, from investors, to our product manager, to our product itself that we just bring the correct KPIs.''} (P14)
Similarly, several developers talked about how running tests slowed them down, but that improving testing and testing infrastructure were not priorities for their companies. 

\textbf{Inability to quantify problems.}
Many developers described how the inability to quantify problems prevented them from making a case for improvement. 
P15 stated: \emph{``It's easy for product to say we have these customers and we can get this revenue, but it's hard to say [invest in improving testing], as there's nothing that clearly illustrates that you're losing 30 collective hours among 30 engineers a week trying to get your testing set up.'' 
}
P10 explained: \emph{``As long as it's not visible in the KPIs, it's not so important for the company. [...] It's hard to make a case for it.''}

\textbf{Lack of visibility/awareness.}
While the lack of measurements can make it difficult to advocate for improvements, developers also struggle with a general lack of awareness about problems both within their teams and with external stakeholders. 
Some developers shared that it was not always clear if colleagues and peers 
experienced similar obstacles. P13 explained: \emph{``I think that definitely parts of this [experience] are shared. I've talked to coworkers. There are definitely some things that they agree upon. I had a difficult time sort of gauging the extent of how my workers feel about the current situation.''} P13 added that colleagues might talk in private about problems but when issues were brought up in meetings, nobody spoke publicly about them. 


\textbf{Lack of buy-in.}
Developers often talked about how the lack of buy-in prevented them from making improvements. Visibility and measurements are important avenues for achieving stakeholder buy-in. In addition, developers described the importance of management trusting and empowering developers.  
P12 explained: \emph{``Good management obviously needs to back [up] these decisions and then support people on that [improving the codebase].''} 

\textbf{Lack of ownership.}
Lack of ownership over friction areas can be a significant barrier. 
For example, P10 described how they encountered major problems in their release pipeline
but that they had minimal capacity to improve it because another team was responsible for it and they did not have permission to make changes.
Having ownership over a problem area can help overcome others barriers---for example, by not having to convince others to work on it or being able to prioritize improvements themselves. 
Previous research has shown that clear ownership increases the willingness of employees to proactively improve their work experiences \cite{Liang2012}. 

\textbf{Undefined expectations.}
Participants often described how they did not think they were expected to make improvements.
This lack of expectations manifests itself individually as well as at the team level. 
P1 shared: \emph{``My perception of normal is that a company posts a job opening that says these are the requirements and the expectations, and then you are expected to come join that team and work the way that team works, as opposed to adding a new person, and then talking about [what can be improved]. This just feels out of the norm to me.''}
Some participants characterized this barrier as the ``acceptance of the status quo''.

\textbf{Lack of incentives.}
The lack of incentives or recognition can discourage improvement efforts. 
For example, P14 talked about poor code reviews and how it was linked to the lack of reward or recognition that P14 and other team members received for doing them. P11 described how they put a lot of effort into improving the test suite, but that the efforts were not seen as valuable as the company was expecting P11 to be a backend Java developer. P11 said they felt undervalued and underappreciated and stopped doing additional work outside of their job description. 

\textbf{Unclear process.}
Several participants explained that having no process for improvement is a barrier, as P6 stated: \emph{``In my last company, the process had a lot of friction and there was nobody improving it. And I was trying to get things done and talk about these things and nothing was happening. So there was no process to improve anything.''}

\textbf{No viable solutions.}
Participants shared that they and their teams sometimes struggled to find solutions to problems. For example, P1 described how their deployment pipeline was cumbersome and causing major productivity losses, but that the team lacked the skills and knowledge to fix it. 
Developers talked about how some problems they experienced were complex organizationally, spanning multiple teams and departments. 
%
Another challenge described by developers was getting different stakeholders to agree on a solution or direction. P13 shared: \emph{``It's difficult to get everyone to agree on one thing and probably almost never happens.'' }

\textbf{Organizational politics.}
Organizational politics and hierarchies were commonly mentioned as standing in the way of improvement. Junior developers shared how being seen as less experienced could reduce their ability to drive change, as others would be more likely to dismiss their problems or solutions. Senior developers also reported struggling with politics. For example, P5 talked about how a management change completely undermined their autonomy and authority, and that power struggles and politics hindered their improvement efforts.

When analyzing our interview data, we found that it is most often not a single barrier that hinders a team to improve.
For example, a lack of ownership over an area can lead to developers not vocalizing a problem within their team because the problem does not seem actionable. This in turn results in a lack of awareness for the organization as a whole, inhibiting others to make improvements. Fortunately, developers also shared with us the strategies they use to overcome these barriers, which we present in the next section.

\subsection{Improvement Strategies}

During our interviews, we inquired about strategies developers and their teams employ to improve developer experience. Developers shared both \emph{individual strategies} and \emph{team strategies}.  

\subsubsection{Individual Strategies}
Individual strategies focus on how developers can change their personal behaviors, environment or activities to drive change.

\textbf{Job crafting.}
One common individual strategy is job crafting, whereby developers actively change their responsibilities or tasks in reaction to poor developer experience. For example, a developer might see that the test suite is not sufficient and will start prioritizing writing automated tests and improving the integration of tests into release and deployment. By doing so, the developer may deviate from their original job description. P11 described: \emph{``I started finding ways to improve our process, or our systems and have a guard rail. I felt that brought value to the business because it would save us from rolling back on releases and reducing any issues [...] I started expanding to things that were getting out of scope.''}

\textbf{Taking risks.}
Many developers talked about how they took risks to improve developer experience, and some mentioned how they would ``ask for forgiveness instead of permission'' when implementing an improvement. P11 stated: \emph{``I know I had influence whenever it came to tools that we could use to help us save time and protect us from new bugs. But, it was really I asked forgiveness instead of permission, there was always that kind of situation. Like no matter what, it would never be a good idea to ask because they would just say no. So you should just do it on the time that you do have, and then be able to show them that like this could bring value, but that also requires you to spend extra time doing that work, that may not be rewarded.''}

\textbf{Speaking up.}
A strategy that every participant mentioned was speaking up. Participants described how speaking up happened during one-on-one meetings, retrospectives, or in casual conversations with colleagues. When developers speak up, they either make problems visible or propose specific improvements. For example, P18 shared: \emph{``I'm just kind of pushy. I started saying: ‘in the past, we haven't had X requirements
and it has led to some rocky development and frustration.' And so I'm not even going to start this ticket until I have these things that I need.''}

And P7 shared: \emph{``In the two retros before [...]  
everybody was supposed to put their smiley that shows their mood [...]
I put it on a three and I put a really angry gorilla face and everybody else did something super casual, like six, seven, happy, smiling, neutral, smiley. But I went really hard and I made a point.''}

\textbf{Local improvements.}
Developers shared how they concentrated on local improvements rather than tackling problems globally. For example, P18 explained that the collaboration process with their design department was problematic. They did not know how to improve the overall processes, but they proactively made arrangements with the designer they worked with: \emph{``Of course I would like to see more global change. The entire company could have a smoother process, but  I'm a junior engineer, so I don't have too much influence. But, I've just taken it upon myself. And so has this designer that's currently working with. We've just taken it upon ourselves to have a more collaborative relationship.''}

\textbf{Workarounds.}
An alternative to making improvements that developers described was coming up with individual workarounds. For example, one developer explained how working on three tickets at the same time helped them deal with slow code review turnaround times and slow test runs: \emph{``It's a very slow test driven development. So, like running a test takes 20 to 40 seconds on the local  machine. Then there is a very long review cycle, which can take up to two weeks to merge a merge request.[...] The way we deal with it is doing everything in parallel. So you work on eight things at a time because you need to switch.''}(P10)

\textbf{Mimicking success.}
Another way to improve developer experience that was described by participants is to mimic the success of others. Developers explained how they looked to others—often more senior members of the team or people that had been in the organization for a longer time—to learn how to navigate and drive change. Observing and mimicking others helped them understand how they could improve developer experience, including the boundaries of individual autonomy and the level of risk-taking allowed. 

\textbf{Being pragmatic.}
Being pragmatic about solutions helps make improvement attainable. Many developers talked about how they balanced their desire to improve developer experience with the goals of their organization. We heard about how developers considered tradeoffs and approached improving developer experience in a sensible and realistic way that created a symbiosis between the developer and the business goals. As P12 explained: \emph{``And it's usually somewhere in between where you end up to be in a good spot, but it's really about having a certain degree of experience how much time you can spend, how you can improve and, and what's necessary and what isn't. So sometimes it's really not so much about making it perfect, but getting it 70 to 80\% there.''}

\subsubsection{Team Strategies}
Team strategies focus on influencing and driving change with the help of others within a team or organization.

\textbf{Building bridges.}
Experienced participants talked about how building bridges was essential for their success in improving developer experience. For example, participants described how they actively and deliberately built relationships with other teams such as product management or quality assurance in order to have ``allies'' that closely understand the developer experience and the presence of various problems.

\textbf{Creating transparency.}
By creating transparency, developers are able to give stakeholders or other teams full visibility into their challenges in order to get the buy-in needed to drive change. P12 explained: \emph{``Everything we do and everything that we define as impacting developer experience is something that the product team also sees because all of these things are not something that happened behind a wall [...] you don't end up in a situation where you have to defend something because they don't understand what's going on or you don't have to push for something that is unrealistic because they don't understand the context, why it's unrealistic. So it's really about having a shared understanding, being able to trust each other, that when I say this is important, my product manager will say, well, then let's do it.''}

\textbf{Convincing others.}
When buy-in from the team or management is needed to make improvements, convincing others about the importance of the problem or potential solutions is an important strategy for successfully driving change. In addition to convincing others, it is important to educate others about how problems may impact product quality or development productivity.

\textbf{Making incremental changes.}
Some problems that affect developer experience are too complex or too large to be solved quickly. Developers talked about how those problems were split up into smaller, more digestible units to be worked on incrementally by the team. For example, reducing technical debt is an issue that many participants described has improved in an incremental manner versus all at once.

\textbf{Having metrics and measurements.}
One strategy for helping engineering teams drive improvement efforts is having a means in place to quantify the problem. Metrics and measurements not only help with making problems visible, but they also help with making progress and improvement efforts tangible. As such, metrics and measurements  also help to reduce the risk that improvement efforts, even if successful, are not visible, and reduce wasted energies on efforts that lead to no improvements. It also helps engineering teams evaluate and learn from their efforts, and build upon their strategies and solution approaches.

\textbf{Having a driver.}
Some participants talked about how having a driver was important in order for improvement efforts to be successful. A driver, as they described, is a person that has specific skills and strengths necessary to champion and execute improvements. Drivers are highly respected within their teams and well connected enough to influence other people whose buy-in is needed. 

\textbf{Involving experts.}
Lastly, participants talked about how involving experts could help in cases where their teams lacked expertise or skills in a certain area. P23 talked about how they lacked knowledge around DevOps pipelines, but they brought in a DevOps engineer to streamline the process from commit to production.

\subsection{Coping Mechanisms}
During the interviews, participants talked about how they cope with poor developer experience that is not sufficiently improved. These coping mechanisms demonstrate concrete ways in which developer experience affects productivity, engagement and retention.

\textbf{Focusing on personal projects.}
One coping mechanism developers described was to focus on personal projects over assigned work. 
Through personal side projects, such as open source work, developers can compensate for missing gratification in their paid work. P8 described: \emph{``During these past six months of where it's been very difficult to ship something and to feel stuck, working on this smaller scale project [outside of work] with these coworkers that I know well has been something that's really helped me to combat the imposter syndrome because, I can see that Hey, when I work with people who I know, and we work on smaller tasks and things that I'm familiar with, I'm able to be productive and to ship code.''}

\textbf{Validating negative experiences.}
Another common coping mechanism was validating negative experiences. The acknowledgment of problems by peers helps people feel better about bad developer experience, even if they cannot improve it---just knowing that others also see and experience those problems is reassuring. P17 shared: \emph{``A lot of it is just affirmation that they hear you. So, when we go to a meeting, we don't necessarily agree with what's happening. When we talk about it afterwards, and it's a lot of, like, I saw that too, or I understand that. 
[...] It's just more acknowledging that we're seeing the same thing and that we don't agree with it, but there's not a whole lot that we can do about it.''} And some developers even talk about how they ``bond over bad developer experience''.

\textbf{Working overtime.}
Poor developer experience can force developers to work overtime to try to make improvements. For example, participants explained that they worked in parallel, did improvement work outside of normal work hours, procrastinated, or took breaks between meetings which led to activities stretching out well beyond the expected work day. 
P13 shared: \emph{``I think that it can definitely lead into that making those [improvement] changes, and doing that work like outside of work hours which I suppose is, like that in itself, isn't great.''}

\textbf{No longer speaking up.}
A commonly described coping mechanism was to stop speaking up about problems, which is the opposite of the strategy to improve their experience. For example, P10 shared: \emph{``But it's like so normal the CI, like everything is slow that it is no longer something that is so much discussed. Because, it also takes more time and we have a specific team that is just doing that. [The team] is also just for basically developer experience or like tooling.''}

\textbf{Reducing engagement.}
Several developers explained how they stopped caring. In those cases, 
they still performed their jobs, but only duties which were absolutely necessary. P8 explained: \emph{``The quality and overall greatness of the software we were building was declining, as we were scaling. And I started to question what was happening? Why are we prioritizing certain projects over improving quality [...] I started to lose trust in like the roadmap and how things were decided to be worked on. And I started being a little bit more vocal and  outspoken about like: 'Hey, we need to care about performance.' [...] but at the end of the day, my arguments were not winning. And, I got really cynical about my experience at [company name] and I started going into work and just being like, whatever, I'm not going to care about these things anymore.''}

\textbf{Gaming the system.}
Another coping mechanism that surfaced during the interviews concerned various ways of gaming the system. For example, one participant explained how he deliberately gave false time estimates that bloated estimated efforts by 100-200\%. This was done to keep expectations from management low and gain time to make improvement efforts that had to be done ``outside'' of normal work hours. As P11 explained: \emph{``[A ticket would] probably take maybe one to three days, but we knew that we don't want to take any more work, so we'll extend it and make it the whole week and just say, `Oh yeah, this is why it took me that long.' I've had some [air quotes] Blockers [air quotes], but they weren't really blockers. It was just more like I'm trying to slow down because I also want to make sure that management doesn't know that I can do my job very well, because then they'll think that is the pace you always need to be at. Or we're going to question your value at this company.''}

\textbf{Leaving their job.}
Nearly every participant brought up ``leaving their job'' as a last resort coping mechanism for dealing with poor developer experience that did not change. P6 shared: \emph{``In my last company, our process had tons of friction all over the place and that led to unhappiness with my job. So it led to, I mean, I left and that was not the whole reason, but that was part of the reason was because getting anything done was hard and it, and it didn't seem like there was, that the company that was focused on improving that.''}

\section{Discussion}
\label{sec:implications}
In this section, we discuss the implications of our research and propose how the \framework\ can be used by both researchers and practitioners.

\subsection{Implications for Researchers}

Developer productivity, motivation, happiness and satisfaction have been active research topics for many years. 
Previous research has surfaced many factors related to these constructs which are relevant to understanding developer experience. 
Our research takes a holistic view of developer experience---building on Fagerholm and M\"{u}nch's research~\cite{fagerholmDeveloperExperienceConcept2012}, we defined developer experience as ``How developers think about, feel about, and value their work'' and the factors we identified address these three aspects. 
 
Our research culminates in the \textbf{DX Framework} which includes DX factors, contextual characteristics impacting factor importance, barriers to improving experience, strategies for improvement, and coping mechanisms of developers when developer experience is not improved (see Fig.~\ref{fig-framework}).
We focused on highlighting \emph{actionable} factors that can pave the way towards identifying and designing interventions to improve developer experience. 
The factors we identified influence all dimensions 
 of the trilogy of mind model (cognition, emotion and conation)~\cite{hilgardTrilogyMindCognition}.
 For example, the development and release factors are closely related to the cognitive dimension of developer experience (i.e., how developers perceive and think about development and release tools), but they also influence developer emotions (e.g., frustration with certain tools) and developer expectations (e.g., potentially less motivation to fix legacy code). 
We also identified contextual characteristics that make specific DX factors more or less \textbf{important} to certain developers. 
An important finding from our research that corroborates previous research~\cite{storeyTheorySoftwareDeveloper2019, murphy-hillWhatPredictsSoftware2019} is that factors related to culture 
and characteristics of the developer's work context matter the most. 
For example, seniority in a team or organization impacts the ``influence''~\cite{farias_what_2019} developers may feel they have to make change to improve their experience while the presence and frequency of problems is also an important determinant of negative experience. 

The \framework\ may be ``put to work''~\cite{varpio2020distinctions} by framing theoretical propositions about how certain interventions may be used to improve developer experience. Although we report on strategies and coping mechanisms shared by our participants, we cannot claim correlation or causal relationships across the constructs captured by our framework. Our hope is that future work will expand our framework and build theories of how to improve developer experience, not just describe or predict it. 
For example, interventions that researchers may wish to study include encouraging and coaching developers to speak up more, to approach their work in smaller increments, and to use the factors we identify to reflect on their experiences. 
Future researchers should also consider developing measurement models for the factors we identified.
For example, some factors may help predict retention, while other factors may be more relevant to predicting developer engagement or quality of the delivered software.

\subsection{Key Takeaways for Industry Practitioners}
In this section, we share key takeaways for industry practitioners as well as propose a three-step process for leveraging the \framework\ to systematically improve developer experience. 

\textbf{Developer experience drives productivity, engagement and job satisfaction.} Our interviews showed that developer experience not only affects a team's ability to get work done, but also developer engagement and the likelihood of developers staying at their current jobs. This highlights the importance for teams and organizations to proactively manage and improve developer experience.
Organization leaders should be aware of the consequences of poor developer experience and work to mitigate the barriers developers and their teams face in trying to improve.

\textbf{Opportunities for improvement are abundant.} All of the developers we interviewed shared at least one actionable factor that had recently affected their developer experience in a negative way. This suggests that opportunities to improve developer experience are widespread and can be surfaced if leaders invite their teams to share and discuss pain points.

\textbf{Factor importance varies, but psychological safety is key.} We discovered that the importance of different factors depends on a developer's role, activities and goals. However, psychological safety was the one exception. Nearly all developers we interviewed described how feeling safe to share opinions and ideas was critical to enabling teams to work together to overcome challenges and implement improvements. Related industry studies (DORA\footnote{A summary of the DORA research can be found at: \url{https://services.google.com/fh/files/misc/dora_research_program.pdf}} and McKinsey\cite{srivastavaDeveloperVelocityHow2020}, Google Aristotle\footnote{A summary of Google Aristotle can be found here: \url{https://rework.withgoogle.com/print/guides/5721312655835136/}}) have similarly identified psychological safety as a top driver of team and business performance.

\textbf{Developer experience is highly personal.} Developer experience is influenced by many factors and moderated by contextual characteristics. This means that developers on the same team can have vastly differing experiences. For example, a senior developer's experience may be significantly affected by poor product management because they are responsible for planning and deadlines. Meanwhile, a junior developer's experience on the same team may not be as affected because they are not concerned with those responsibilities. Because developer experience is highly personal, achieving good developer experience often involves tradeoffs and compromises. For example, P4 shared how doing code reviews for their team members hurt their own experience (because they would rather spend time working on features) while benefitting the team. P4 added that they deliberately "sacrifice" their experience for the greater good of the team.

\textbf{Developer experience involves both organizational and technical challenges.} We found through our interviews that the barriers impeding a team's ability to improve developer experience are both organizational and technical. As described earlier, psychological safety is paramount for enabling developers to speak up about problems and share solutions for improving developer experience.

\textbf{The DX Framework can be put to work using an Ask-Plan-Act process.}
We anticipate that the \framework\ may be used as a foundation for systematic approaches to assessing and improving developer experience. To do so, we suggest a three-step process outlined below. This proposed methodology is based on our interview findings, the expertise of our research team and discussions with practitioners. 

\begin{enumerate}
\item \emph{Ask: Make problems visible.}
The first step to assessing developer experience is to learn about the day-to-day experiences of developers. As factor importance is highly dependent on the responsibilities, activities and goals of each developer, it is crucial to ask each individual developer about their experience. Our factors and barriers can be used as prompts for collecting feedback through structured feedback mechanisms (e.g., surveys) or unstructured methods (e.g., retrospectives, one-on-one meetings). 

\item \emph{Plan: Improvement needs to happen on an individual, team and organizational level.}
Once developer feedback has been collected, it can be analyzed to determine the areas that need to be improved. 
To ensure that individuals and teams are empowered to make improvements, we recommend assigning explicit owners for each improvement area. Improvement efforts need to be planned for and be allocated appropriate time and resources. 

\item \emph{Act: Continuous, small improvements are key.}
To work on improvements, developers and teams may benefit from working incrementally with fast feedback loops. Strategies from the \framework, such as \emph{building bridges}, \emph{convincing others} and \emph{creating transparency}, may help overcome barriers to improvement. 
Developers can use individual strategies to make improvements themselves without relying solely on their teams or leaders to drive change.
The factors can also be used to derive measures for assessing developer experience and monitoring the success of improvement efforts. 
This makes both the problems and improvement efforts more visible.
Once planned improvements have been implemented, this three-step process can be repeated to continue progress on a continuous basis.
\end{enumerate}


\section{Threats to Credibility and Reliaibility}
\label{sec:threats}

In contrast to quantitative studies, qualitative studies are more prone to threats to credibility than threats to validity. Validity and reliability in qualitative work mostly has to do with how careful, thorough and honest the researchers have been during data collection and analysis~\cite{robson_real_2002}.  Therefore, in the following, we mainly describe threats to external and internal credibility of our study.

To increase our thoroughness and trustworthiness, we developed an interview guideline which was improved after two pilot interviews, and we thoroughly coded each transcript (which was automatically transcribed) iteratively. As transcripts were directly linked to the applicable video recording of each participant, the researchers could make sure that any errors introduced by automatic transcription were corrected. During coding, the researcher also frequently replayed a video to ensure a clear understanding of what a segment of transcript was about and to increase correct interpretation of the meaning, through not only reading the transcript, but also by hearing (tone) and seeing (body language) the participants. This helped us ensure that we understood the context and content of the statement as much as possible.

Another threat to internal credibility for our study is interpretive validity, which describes the threat that the researchers imposed their own framework or meaning rather than understanding the perspectives of the participants and the meaning their words and explanations had~\cite{maxwell_understanding_1992}.  We mitigated this threat by paraphrasing many of the key statements made during the interviews and asked clarifying questions. In addition, while the main coding was done by the first author of this study, the other two authors were extensively involved with the axial coding process and the establishment of the emerging factors, barriers, coping mechanisms and categories. 
In addition, we also kept an extensive audit trail in the form of recorded videos and complete transcripts from all participants. All coding steps were documented and available to all researchers. Parts of this is also available as supplemental data.

With respect to external credibility, our sample size of 21 participants exceeds what Guest et al. \cite{guestHowManyInterviews2006} recommends for achieving saturation as our group of participants were a relatively homogeneous group of active software developers. In addition, as reported, no new categories or concepts emerged during the last three interviews, which makes us confident that saturation was reached. 

We recognize that how we framed the initial questions in the interviews, and how we informally defined the concept of developer experience and how we asked them to describe ``factors'' that impact their experience may have influenced how participants responded in their interviews and in terms the reliability of our study. 
Any alternative wording would have suffered from similar challenges.  By having follow up questions, we tried to mitigate this to some extent.  
We also prompted the participants to think about their teamwork as we (and others, such as Fagerholm and M\"{u}nch~\cite{fagerholmDeveloperExperienceConcept2012}) consider teamwork as an important concern to discuss. Of course, by prompting them to think of their teams, we may have biased their responses. 

During the interviews, we showed participants a list of factors that may impact their experience (as guided by a review of the previous research) to systematically steer a discussion around importance and actionability. While this was helpful to conquer recency bias and help participants consider more factors, it also introduces the threat of confirmation bias and respondent bias. For example, participants might confirm factors in order to not offend the researcher. To mitigate this threat, we only considered factors that came up before the additional factor list was shown in our analysis to distill the DX factors. In addition, we asked participants during the interviews to report what they experienced and to share details of those experiences. In these cases, it was clear that we analyzed data that represents a participant's experience and not what they thought the researcher might want to hear.

\section{Conclusion}
\label{sec:conclusion}
In this paper, we presented an actionable conceptual framework that identifies the main factors that affect developer experience. Our research shows that a factor's level of importance depends on characteristics that shape the context in which developers experience those factors. 
In addition, we identified barriers that prevent development teams from improving their experience as well as strategies for overcoming those barriers. Lastly, we discussed coping mechanisms of developers when developer experiences is not sufficiently improved.  
By improving developer experience, organizations can improve developer productivity, satisfaction, retention, and organizational performance. Thus, developer experience is key to helping both developers and businesses thrive. Our framework provides a go-to reference to help organizations understand what is important to create a productive, effective and satisfying environment for developers, and points to future research for understanding, measuring and improving developer experience.
We conclude with a proposal for a summarized definition for developer experience that is an aggregation of many comments we heard throughout our interviews. This new definition, which also nicely captures the trilogy of mind concepts of cognition, conation and affect, is something all teams should strive for: 

\begin{center}
\emph{Being empowered to do my best work, joyfully.}
\end{center}


%



\ifCLASSOPTIONcompsoc
  \section*{Acknowledgments}
\else
  \section*{Acknowledgment}
\fi

We want to thank Alberto Bacchelli and Cassandra Petrachenko for their valuable input and guidance on our paper.
We also thank the interview participants for sharing with us their perspectives and strategies for improving developer experience.

\ifCLASSOPTIONcaptionsoff
  \newpage
\fi



\bibliographystyle{IEEEtran}
\bibliography{Dx-bib}
\end{document}